# Influence of Mutual Drag of Light and Heavy Holes on conductivity of $p$-Silicon and $p$-Germanium

## I.I. Boiko[1]

*Institute of Semiconductor Physics, NAS of Ukraine, 45 prospect Nauky, 03028 Kyiv, Ukraine*
(Dated: November 22, 2010)

Conductivity of $p$-Si and $p$-Ge is considered for two band model with due regard for mutual drag of light and heavy holes. It is shown that for small and moderate temperatures this drag significantly diminishes drift velocity of light holes and, as result, the whole conductivity of crystal. Considered here drag effect appears as well in the form of nonmonotonous dependences of conductivity on temperature and carrier concentration

## 1. Introduction

. In one of previous work we have investigated the direct influence of interelectron interaction on conductivity of $n$-silicon [1]. In contrast to crystals with one simple band, where the electron-electron scattering does not change total momentum of carriers, in multyvalley crystals the conductivity can be essentially influenced by mutual drag of carriers, which belong to different valleys (see Refs. [1, 2]). The analogous effect has to appear in semiconductors where band carriers occupy several subbands and transitions between these subbands are sufficiently rare. If mobilities of carriers from separate subbands are appreciably different, then values of separate drift flows are determined not only by external scattering system (phonons, impurities) but by mutual drag of band carriers also. In this case more quick flow of carriers is inhibited by more slow flow, and the latter is accelerated by the first. This drag results in a change of total conductivity of crystal.

The convenient objects for investigation of mutual drag are $p$-silicon and $p$-germanium. We consider here two band model for this crystals, and for simplicity of calculations accept spherical bands approximation. So dispersion law has the parabolic form:

$$\varepsilon_{\vec{k}}^{(a)} = \hbar^2 k^2 / 2m_a \quad . \quad (a = 1 \text{ or } 2) \tag{1}$$

Here $m_a$ are effective masses ($m_1$ is mass of light holes and $m_2$ is mass of heavy holes).

Previous calculations show that approximation of isotropic parabolic bands introduces in calculated conductivity an inaccuracy about several percents only. Mutual drag, as it will be shown below, can change conductivity much more.

---

[1] E-mail: igorboiko@yandex.ru

In the case $m_1 << m_2$ the concentration of light holes $p_1$ differs substantially from concentration of heavy holes $p_2$ (see Ref. [3]). We have $p_2 / p_1 = (m_2 / m_1)^{3/2} = (0.33 m_0 / 0.04 m_0)^{3/2} =$ $= 23.7$ for $p$-germanium and $p_2 / p_1 = (m_2 / m_1)^{3/2} = (0.56 m_0 / 0.16 m_0)^{3/2} = 6.55$ for $p$-silicon. One can see that concentration of light holes is small, and they cannot drag heavy holes noticeable. Therefore heavy holes are not sensitive to the drag by light holes. Quite opposite situation we have for light holes. In spite of small number their contribution in total conductivity is quite comparable with that of heavy holes. Therefore drag of light holes by heavy holes can influence essentially on total conductivity.

## 2. **Balance equations**

Many years ago (see Ref. [3]) some special attempt was taken to consider influence of intervalley scattering on cyclotron resonance in silicon. Performing the calculations authors of noted article followed directions of Ref. [4], where some additional scattering term for kinetic equation (in the form of tau-approximation) was proposed. It was not good idea, because collision integral for $e$-$e$-scattering principally cannot be represented in a form containing some relaxation time (see Ref. [5]). We use here quite another approach, which allows to involve in consideration interaction of band carriers with sufficiently good reasons. (see, for example, Refs. [6] and [7]).

Consideration of conductivity we begin from the set of two balance equations, obtained as a first momentum of quantum kinetic equations (see Refs [2, 7] ):

$$e \, \vec{E} + \vec{F}^{(a)} + \sum_{b=1}^{2} \vec{F}^{(a,b)} = 0 \, , \qquad (a = 1, 2) \qquad (2)$$

These equations contain electrical force $e\vec{E}$ as dynamic that, and forces, generated by scattering system, as friction. The resistant force related to external scattering system is

$$\vec{F}^{(a)} = -\frac{e^2}{(2\pi)^6 \, p_a} \int d^3\vec{k} \int \vec{q} \, d^3\vec{q} \int d\omega \, \delta(\hbar\omega - \varepsilon_{\vec{k}}^{(a)} + \varepsilon_{\vec{k}-\vec{q}}^{(a)}) \{ f_{\vec{k}}^{(a)} - f_{\vec{k}-\vec{q}}^{(a)} + [ f_{\vec{k}}^{(a)} (1 - f_{\vec{k}-\vec{q}}^{(a)}) +$$
$$+ f_{\vec{k}-\vec{q}}^{(a)} (1 - f_{\vec{k}}^{(a)}) ] \tanh(\hbar\omega / 2 k_B T) \} [ \langle \varphi^2_{(I)} \rangle_{\omega, \vec{q}} + \langle \varphi^2_{(ph)} \rangle_{\omega, \vec{q}} \, ] \, . \qquad (3)$$

We consider here interaction of holes with acoustic phonons and charged impurities, disposed uniformly in space. In Eq. (3) the value $f_{\vec{k}}^{(a)}$ is nonequilibrium distribution function for $a$-holes ; the values $\langle \varphi^2_{(I)} \rangle_{\omega, \vec{q}}$ and $\langle \varphi^2_{(ph)} \rangle_{\omega, \vec{q}}$ are Fourier-images for correlators of charged impurities and acoustic phonons scattering potentials. In our calculations we use such forms (see Refs. [2], [7]):

$$\langle \varphi^2_{(I)} \rangle_{\vec{q},\omega} = \frac{32\pi^3 e^2 n_I}{\varepsilon_L^2 (q^2 + q_0^2)^2} \delta(\omega) ; \tag{4}$$

$$\langle \delta\varphi^2_{(ph)} \rangle_{\vec{q},\omega} = \Xi_A^2 \frac{2\pi k_B T}{e^2 \rho \ s^2} \delta(\omega) . \tag{5}$$

Here $n_I$ is concentration of charged impurities, $\varepsilon_L$ is dielectric constant of crystal lattice, $\Xi_A$ is deformation potential constant. The form (5) corresponds to the approximation of quasielastic collisions. For nondegenerate carriers

$$q_0^2 = \frac{4\pi e^2 (p_1 + p_2)}{\varepsilon_L k_B T} . \tag{6}$$

The Coulomb interaction between holes is presented by the forces ($a$ , $b = 1, 2$)

$$\vec{F}^{(a,b)} = \frac{e^4 \hbar}{4\pi^6 p_a} \int \vec{k} \ d^3\vec{k} \int d^3\vec{k}' \int d^3\vec{q} \ \frac{1}{q^4} \frac{\delta(\varepsilon_{\vec{k}}^{(a)} - \varepsilon_{\vec{k}-\vec{q}}^{(a)} - \varepsilon_{\vec{k}'}^{(b)} + \varepsilon_{\vec{k}'-\vec{q}'}^{(b)})}{\varepsilon_L^2 (q^2 + q_0^2)^2} G_{ab}(\vec{k},\vec{k}',\vec{q}) ;$$

$$G_{ab}(\vec{k},\vec{k}',\vec{q}) = f_{\vec{k}-\vec{q}}^{(a)} (1 - f_{\vec{k}}^{(a)}) f_{\vec{k}'}^{(b)} (1 - f_{\vec{k}'-\vec{q}}^{(b)}) - f_{\vec{k}}^{(a)} (1 - f_{\vec{k}-\vec{q}}^{(a)}) f_{\vec{k}'-\vec{q}}^{(b)} (1 - f_{\vec{k}'}^{(b)}) . \tag{7}$$

To calculate the drift velocities $\vec{u}^{(a)}$ of holes from $a$-group we accept the model of non-equilibrium distribution functions as Fermi functions with argument containing shift of velocity $\vec{v}^{(a)}(\vec{k}) = \hbar^{-1} (\partial \varepsilon_{\vec{k}}^{(a)} / \partial \vec{k})$ on correspondent velocity $\vec{u}^{(a)}$ :

$$f_{\vec{k}}^{(a)} = f^{0(a)}(\vec{v}^{(a)}(\vec{k}) - \vec{u}^{(a)}) . \quad (a = 1, 2) \tag{8}$$

Here $f^{0(a)}(\vec{v}^{(a)}(\vec{k})) = f_0^{(a)}(\varepsilon)$ is equilibrium distribution function for $a$-carriers. Drift velocities $\vec{u}^{(a)}$ are proportional to partial densities of currents $\vec{j}^{(a)}$ :

$$\vec{u}^{(a)} = \frac{1}{e p_a} \vec{j}^{(a)} . \tag{9}$$

The density of total current is

$$\vec{j} = \sum_{a=1}^{2} \vec{j}^{(a)} . \tag{10}$$

Using the forms (8) and carrying out linearization of forces in Eqs. 2 over drift velocities we obtain due to spherical symmetry the following simple set of balance equations:

$$\vec{F}^{(a)} = -e \ \beta^{(a)} \ \vec{u}^{(a)} ; \quad \vec{F}^{(a,b)} = -e \ \xi^{(a,b)} (\vec{u}^{(a)} - \vec{u}^{(b)}) . \quad (a , b = 1, 2) \tag{11}$$

Here coefficients $\beta^{(a)}$ and $\xi^{(a,b)}$ are (see Refs [2, 7]):

$$\beta^{(a)} = \lambda^{(a)} + \chi^{(a)} \ ; \tag{12}$$

$$\lambda^{(a)} = \frac{\hbar}{6(2\pi)^5 e \ p^{(a)} k_B T} \int d\omega \int \frac{q^2 \, d^3\vec{q}}{\sinh(\hbar\omega/k_B T)} B_{(a)}(\omega,\vec{q}) \langle \varphi^2_{(I)} \rangle_{\omega,\vec{q}} =$$

$$= \frac{e \, n_I}{6\pi^2 \ p^{(a)} \varepsilon_L^2} \int \left[ \frac{1}{\omega} B_{(a)}(\omega,\vec{q}) \right]_{\omega=0} \frac{q^2}{(q^2+q_0^2)^2} d^3\vec{q} \ ; \tag{13}$$

$$\chi^{(a)} = \frac{\hbar}{6(2\pi)^5 e \ p^{(a)} k_B T} \int d\omega \int \frac{q^2 \, d^3\vec{q}}{\sinh(\hbar\omega/k_B T)} B_{(a)}(\omega,\vec{q}) \langle \varphi^2_{(ph)} \rangle_{\omega,\hat{q}} =$$

$$= \frac{\Xi_A^2 \, k_B T}{96\pi^4 p^{(a)} e^3 \rho \, s^2} \int \left[ \frac{1}{\omega} B_{(a)}(\omega,\vec{q}) \right]_{\omega=0} q^2 \, d^3\vec{q} \ ; \tag{14}$$

$$\xi^{(a,b)} = \frac{\hbar^2 (1-\delta_{ab})}{6(2\pi)^4 e \ \varepsilon_L^2 p^{(a)} k_B T} \int \frac{d\omega}{\sinh^2(\hbar\omega/2k_B T)} \int \frac{q^2 \, d^3\vec{q}}{(q^2+q_0^2)^2} B_{(a)}(\omega,\vec{q}) B_{(b)}(\omega,\vec{q}) \ . \tag{15}$$

Here

$$B_{(a)}(\omega \ ,\vec{q}) = -\frac{e^2}{\pi} \int d^3\vec{k} [f_0(\varepsilon^{(a)}_{\vec{k}}) - f_0(\varepsilon^{(a)}_{\vec{k}-\vec{q}})] \, \delta(\varepsilon^{(a)}_{\vec{k}-\vec{q}} - \varepsilon^{(a)}_{\vec{k}} + \hbar\omega) \ . \tag{16}$$

For quasielastic collisions we have the form (see Ref. [2])

$$B_{(a)}(\omega \to 0, \vec{q}) = \omega \Lambda^{(a)}(\vec{q}) \ ,$$

where

$$\Lambda^{(a)}(\vec{q}) = \frac{2e^2 \, m_a^2}{q \, \hbar^3} \left[ 1 + \exp\left\{ \frac{\hbar^2 q^2}{8 k_B T \, m_a} - \eta \right\} \right]^{-1} . \tag{17}$$

Here $\eta$ is dimensionless Hermi-energy:

$$\eta = \varepsilon_F / k_B T \ . \tag{18}$$

As result we have for nondegenerate holes:

$$\lambda^{(a)} = \frac{4\sqrt{2\pi m_a} e^3 \, n_I}{3(k_B T)^{3/2} \varepsilon_L^2} \int_0^\infty \frac{q^3 dq}{(q^2+q_0^2)^2} \exp\left(-\frac{\hbar^2 q^2}{8 m_a k_B T}\right) \ ; \tag{19}$$

$$\chi^{(a)} = \frac{8\sqrt{2} \, \Xi_A^2 (k_B T)^{3/2} m_a^{5/2}}{3\pi^{3/2} \hbar^4 e \rho \ s^2} \ ; \tag{20}$$

$$\xi^{(a,b)} = \frac{8\gamma \, e^3 \, m_b^2 \, p_a}{3 k_B T \, \hbar \, m_a} \int_0^\infty \frac{q^2 dq}{(q^2+q_0^2)^2} \exp\left[-\frac{\hbar^2 q^2}{8 k_B T}\left(\frac{1}{m_1}+\frac{1}{m_2}\right)\right], \tag{21}$$

where

$$\gamma = \int\limits_{-\infty}^{\infty} \frac{w^2 dw}{\sinh^2 w} \approx 3.29 \cdot \tag{22}$$

Note, that

$$p_a \xi^{(a,b)} = p_b \xi^{(b,a)}; \tag{23}$$

Therefore $\xi^{(2,1)} = (p_1 / p_2)\xi^{(1,2)} \equiv w\xi^{(1,2)}$. Farther we use the designation $\xi^{(1,2)} = \xi$. For germanium $w = 0.042$, for silicon $w = 0.153$ (see Ref. [8]). We also assume $n_l = p = p_1 + p_2$.

As result we have the following system for drift velocities of light and heavy holes:

$$\vec{E} = \beta^{(1)}\vec{u}_1 + \xi(\vec{u}_1 - \vec{u}_2) \quad ; \quad \vec{E} = \beta^{(2)}\vec{u}_2 + w\xi(\vec{u}_2 - \vec{u}_1). \tag{24}$$

The case $\xi = 0$ corresponds to neglect of mutual drag of light and heavy holes.

## 3. Mobility of holes

Solving the system (24) one obtains the following expressions for drift velocities of light and heavy holes:

$$\vec{u}_1(\xi) = \frac{\beta^{(2)} + (w+1)\xi}{\beta^{(1)}\beta^{(2)} + \xi(\beta^{(2)} + w\beta^{(1)})}\vec{E} \equiv \mu_1(\xi)\vec{E} \quad ;$$

$$\vec{u}_2(\xi) = \frac{\beta^{(1)} + (w+1)\xi}{\beta^{(1)}\beta^{(2)} + \xi(\beta^{(2)} + w\beta^{(1)})}\vec{E} \equiv \mu_2(\xi)\vec{E} \quad . \tag{25}$$

One finds from here the dependences of drift velocities on drag coefficient $\xi$. For relative drift values of dragged holes

$$\frac{u_1(\xi)}{u_1(0)} = \frac{\beta^{(1)}[\beta^{(2)} + (w+1)\xi]}{\beta^{(1)}\beta^{(2)} + \xi(\beta^{(2)} + w\beta^{(1)})} \quad ; \quad \frac{u_2(\xi)}{u_2(0)} = \frac{\beta^{(2)}[\beta^{(1)} + (w+1)\xi]}{\beta^{(1)}\beta^{(2)} + \xi(\beta^{(2)} + w\beta^{(1)})} \quad . \tag{26}$$

Introduce the total conductivity $\sigma(\xi)$ and the average hole mobility $\mu(\xi)$ with the help of following relations:

$$\sigma(\xi) = \frac{e}{E}[p_1 u_1(\xi) + p_2 u_2(\xi)] \quad ; \quad \mu(\xi) = \frac{\sigma(\xi)}{p_1 + p_2} = \frac{\sigma(\xi)}{p} \quad . \tag{27}$$

## 4. Results of numerical calculations

In this article our numerical calculations are performed for nondegenerate holes. Figure 1 shows areas, substantially different in relation to degeneracy. Presented there separating lines correspond to the case $\varepsilon_F = 0$.

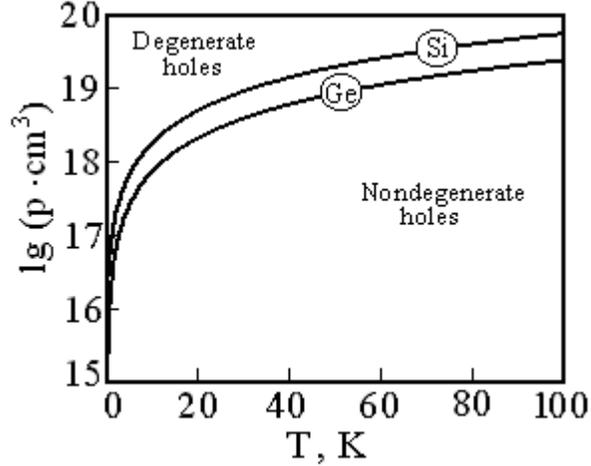

Fig. 1. Areas of degenerate and nondegenerate holes.

To perform numerical calculations we use the following values:

$\varepsilon_L = 12$, $\rho s^2 = 1.66 \cdot 10^{11} Pa$, $\Xi_A = -4.2\,eV$ for $p$-silicon and

$\varepsilon_L = 16$, $\rho s^2 = 1.26 \cdot 10^{11} Pa$, $\Xi_A = 1.9\,eV$ for $p$-germanium.

Fig. 2 shows the ratio of drift velocities $u_1$, calculated for drugged (there we have $u_1(\xi)$) and for undrugged light holes (there we have $u_1(0)$). It is evident that drug by heavy holes significantly diminishes (in several times) the drift velocity of light holes.

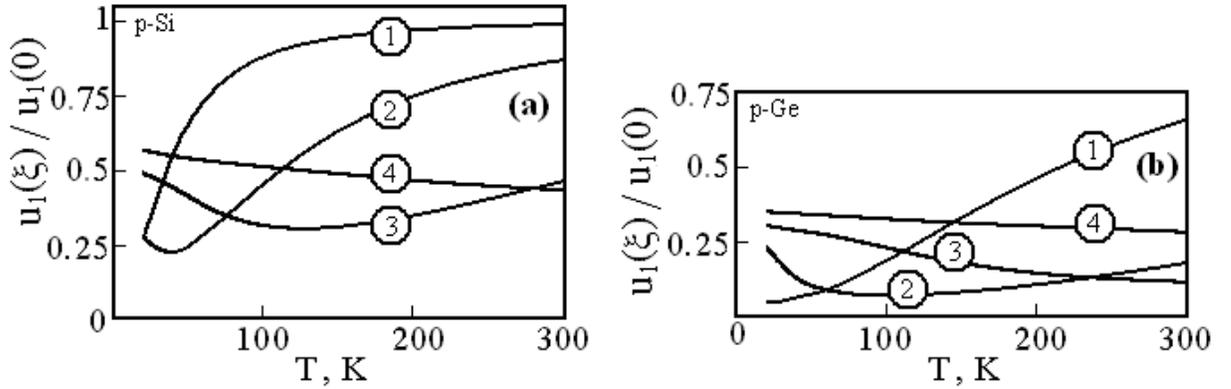

Fig. 2 (a, b).Dependence of relative drift velocity of light holes on temperature.

$1 - p = 10^{12}\,cm^{-3}$ ; $2 - p = 10^{14}\,cm^{-3}$ ; $3 - p = 10^{16}\,cm^{-3}$ ; $4 - p = 10^{18}\,cm^{-3}$ .

Fig. 3 (a, b) allows to compare visually the drift velocities calculated for drugged and undrugged heavy holes. One can see that drug by light holes increases the drift velocity of heavy holes only on a little percents.

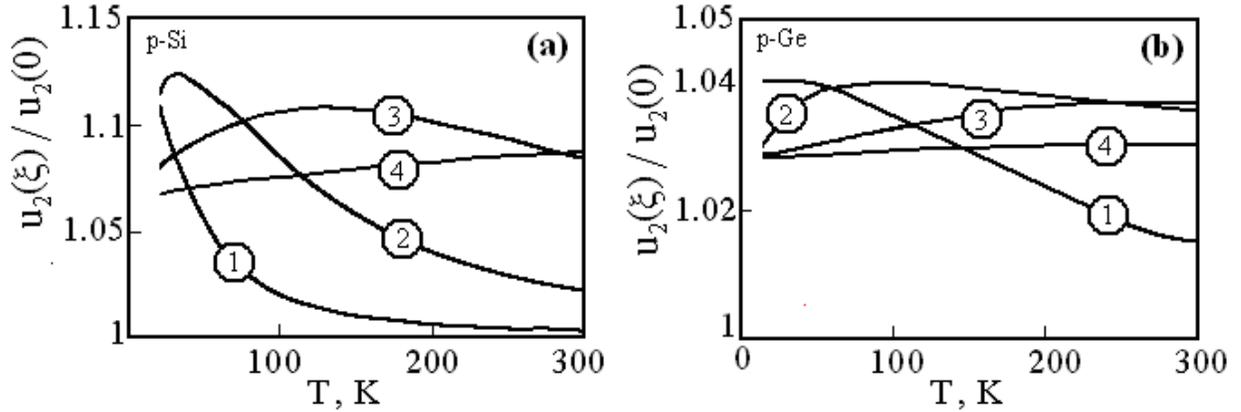

Fig. 3 (a, b). Dependence of relative drift velocity of heavy holes on temperature.

$1 - p = 10^{12}\,cm^{-3}$ ; $2 - p = 10^{14}\,cm^{-3}$ ; $3 - p = 10^{16}\,cm^{-3}$ ; $4 - p = 10^{18}\,cm^{-3}$ .

Figs. 4 (a, b) and 5 (a, b) demonstrate level of influence of mutual drag of light and heavy holes on average mobility of $p$-silicon and $p$-germanium. The main result of carried calculations is the absolute decrease of the mobility due to mutual drag. It should point out the complicated dependences of the ratio $\mu(\xi)/\mu(0)$ on temperature and on total concentration of holes.

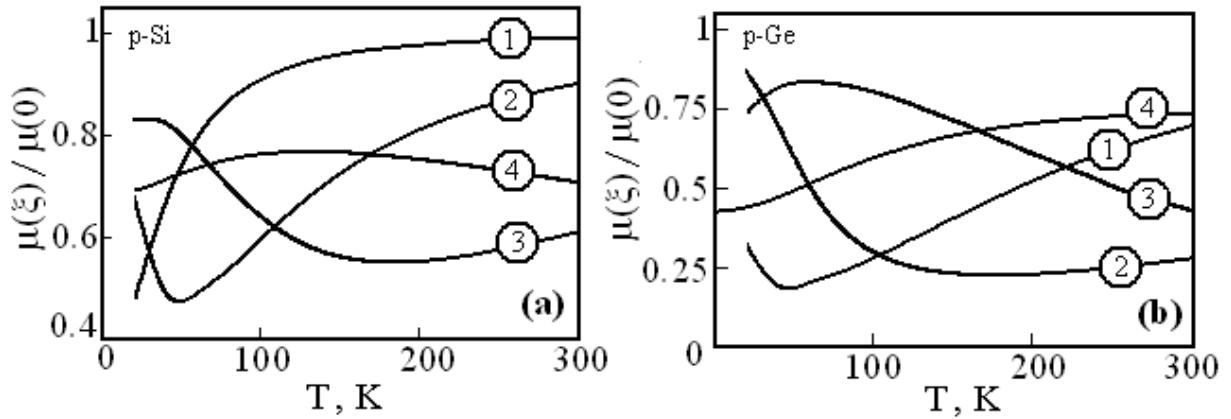

Fig. 4 (a, b). Dependence of relative mobility on temperature.

$1 - p = 10^{12}\,cm^{-3}$ ; $2 - p = 10^{14}\,cm^{-3}$ ; $3 - p = 10^{16}\,cm^{-3}$ ; $4 - p = 10^{18}\,cm^{-3}$ .

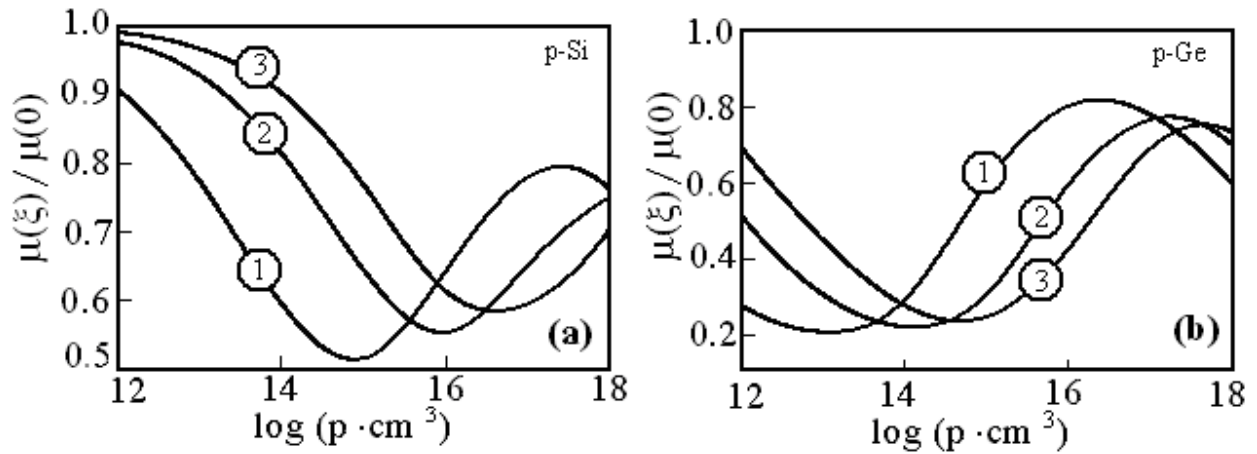

Fig. 5 (a, b). Dependence of relative mobility on hole concentration.

$1 - T = 100 \, K, \quad 2 - T = 200 \, K, \quad 3 - T = 300 \, K$.

---